\documentclass[twoside,11pt]{article}

\usepackage[preprint]{jmlr2e}
\usepackage{amsmath}
\usepackage{amssymb, mathtools, amsfonts}
\usepackage{tikz}
\usepackage{tikz-qtree}
\usetikzlibrary{trees}
\usepackage{algorithm}
\usepackage[frozencache, cachedir=minted-cache]{minted} % First pip install pygments
\usepackage[capitalize]{cleveref}
\usepackage{subcaption}
\usepackage{wrapfig}
\usepackage{ragged2e}

\setminted[python]{framesep=2mm, fontsize=\footnotesize, numbersep=5pt}

\newcommand{\G}{\mathcal{G}}
\newcommand\ci{\perp\!\!\!\perp}
\DeclareMathOperator{\diedgeright}{\textcolor{blue}{\boldsymbol{\rightarrow}}}

\DeclareMathOperator{\biedge}{\textcolor{red}{\boldsymbol{\leftrightarrow}}}
\DeclareMathOperator{\udedge}{\textcolor{brown}{\boldsymbol{--}}}

\jmlrheading{}{2022}{}{08/22}{??/??}{lee-bhattacharya22}{Lee, Bhattacharya, Nabi, and Shpitser}

\ShortHeadings{Ananke}{Lee, Bhattacharya, Nabi, and Shpitser}
\firstpageno{1}

\begin{document}

\title{\texttt{Ananke}: A Python Package For Causal Inference Using Graphical Models}
\author{\name Jaron J. R. Lee$^{\dagger, 1}$ \email jaron.lee@jhu.edu 
	%Baltimore, MD 21218, USA
    \AND
    \name Rohit Bhattacharya$^{\dagger, 2}$ \email rb17@williams.edu 
	%Williamstown, MA 01267, USA
	\AND
	\name Razieh Nabi$^{3}$ \email razieh.nabi@emory.edu 
	%Atlanta, GA 30322, USA
	\AND
	\name Ilya Shpitser$^{1}$ \email ilyas@cs.jhu.edu 
	%Baltimore, MD 21218, USA
	\AND
	\name $\dagger$ Equal contribution\\
    \addr $^1$Department of Computer Science, Johns Hopkins University \\%
	\addr $^2$Department of Computer Science, Williams College\\
    \addr $^3$Department of Biostatistics and Bioinformatics, Emory University 
    }
\editor{}

\maketitle

\begin{abstract}%   <- trailing '%' for backward compatibility of .sty file
	
 We implement \texttt{Ananke}: an object-oriented Python package for causal inference  with graphical models. At the top of our inheritance structure is an easily extensible \texttt{Graph} class that provides an interface to several broadly useful graph-based algorithms and methods for visualization. We use best practices of object-oriented programming to implement subclasses of the \texttt{Graph} superclass that correspond to  types of causal graphs that are popular in the current literature. This includes directed acyclic graphs for modeling causally sufficient systems, acyclic directed mixed graphs for modeling unmeasured confounding, and chain graphs for modeling data dependence and interference. 
 Within these subclasses, we implement specialized algorithms for common statistical and causal modeling tasks, such as separation criteria for reading conditional independence, nonparametric identification, and parametric and semiparametric estimation of model parameters. Here, we present a broad overview of the package and example usage for a problem with unmeasured confounding. Up to date documentation is available at \textcolor{blue}{\url{https://ananke.readthedocs.io/en/latest/}}.
\end{abstract}


\section{Introduction}

Causal inference is a pipeline comprised of many steps -- specification of a causal model, identification of the desired causal parameter under assumptions of this model, estimation of the parameter from data based on the identifying functional, and robustness checks via sensitivity analysis and uncertainty quantification. Any of these steps may be complicated by unmeasured confounding, data dependence, and missing data. In \texttt{Ananke}, we implement methods that span all of these steps, including nonparametric identification and semiparametric estimation strategies, to provide analysts  a unifying interface that allows them to set up end-to-end pipelines that exhibit robustness to the aforementioned complications. %, and often, an analyst will have to revisit steps in the pipeline more than once.

In particular, we adopt an object-oriented paradigm to implement graph-based causal inference methods. We build an inheritance structure spanning causal graphical models that use any combination of directed ($\diedgeright$), bidirected ($\biedge$), and undirected ($\udedge$) edges. We hope that due to its object-oriented nature and easily accessible Python implementation \texttt{Ananke} will improve the accessibility of many graph-based causal inference methods, and allow interested users to easily extend and build on its current infrastructure.

\textbf{Related work}: The \texttt{doWhy} package and \texttt{DAGitty} aim to provide a unifying interface for distinct steps in the causal inference pipeline. However, their estimation capabilities are largely limited to settings without unmeasured confounders or selection bias. In the case of \texttt{doWhy}, the causal graph  interface is still under active development with plans to interface with \texttt{Ananke} rather than build one from scratch (personal communication with developers.) Other existing packages emphasize a single step in the pipeline. \texttt{TETRAD} and its Python port \texttt{causal-learn} \citep{scheines1998tetrad}, \texttt{pcalg} \citep{kalisch2012causal}, and \texttt{cdt} \citep{kalainathan2020causal} focus on graph representation and model selection; \texttt{causaleffect} focuses on nonparametric identification; \texttt{npcausal} \citep{kennedy2021npcausal}, \texttt{zEpid} \citep{zevich2018zepid}, \texttt{tmle3} \citep{coyle2021tmle3-rpkg}, and \texttt{DoubleML} \citep{DoubleML2022Python} focus on semiparametric estimation. %The \texttt{doWhy} package \citep{dowhy} aims to provide a unifying interface in Python for the causal inference pipeline, although their graphical causal model interface is still under active development. \texttt{tmle3}\citep{coyle2021tmle3-rpkg} is an R package wrapping around existing packages to provide abstractions (e.g. graphs, parameters) for use in a general pipeline. Other packages have tended to focus on part of the pipeline: for graphical representation and model selection, \texttt{TETRAD}\citep{scheines1998tetrad} and its Python port \texttt{causal-learn} has been the gold standard for many years; for identification, 
%\texttt{causaleffect} \citep{tikkaIdentifyingCausalEffects2017} provides algorithms for identification under unmeasured confounding and missing data, but lacks estimation capabilities; for estimation, packages like \texttt{DoubleML} \citep{DoubleML2022Python} provide estimation strategies under no unmeasured confounding.
Other standalone packages exist as appendices to papers, and in certain cases we reimplement these algorithms in \texttt{Ananke}, e.g., the maximum likelihood algorithms in \cite{evans2013admgs} and \cite{drton2009computing}.
The principle advantage of \texttt{Ananke} over peers is that it offers a unified and easily extended interface for causal inference in a single package, with an active community. %In \texttt{ananke}, we aim to complement these pieces by providing additional identification strategies, and estimation procedures that are not implemented in \texttt{doWhy}. 

\section{Overview of \texttt{Ananke}'s Graph Inheritance Structure}
\label{sec:hierarchy}

\begin{wrapfigure}[12]{r}{0.48\textwidth}%[lineheight]{position}{width}
    \vspace{-0.5cm}
    \centering
    \includegraphics[width=0.48\textwidth]{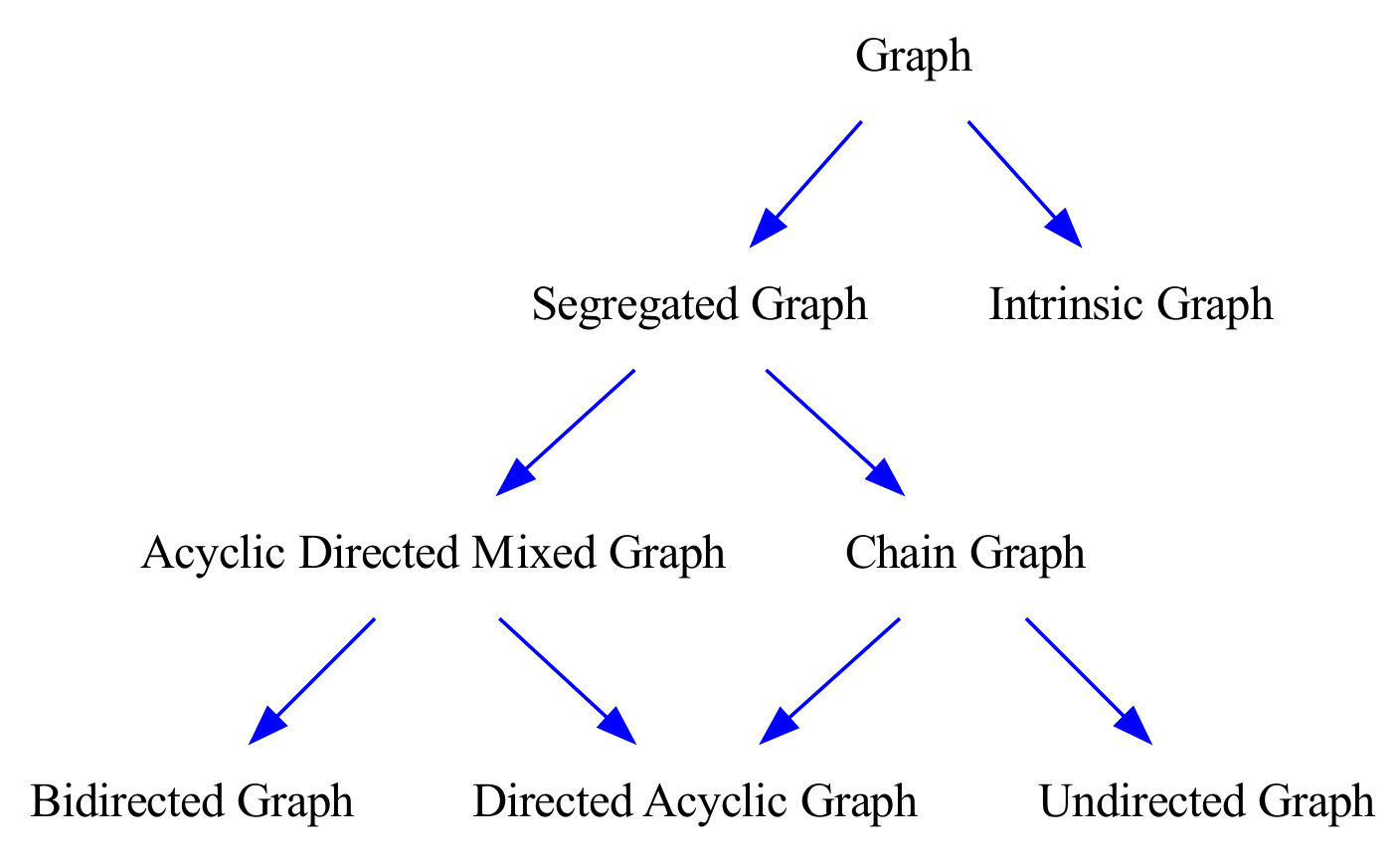}
    \caption{\small \texttt{Ananke}'s graph inheritance structure.}
    \label{fig:hierarchy}
\end{wrapfigure}
An overview of graphical models in \texttt{Ananke} and their inheritance structure is shown in Fig.~\ref{fig:hierarchy}. The \texttt{Graph} class currently supports the creation of graphs $\G = (V, D, B, U)$, where $V$ denotes a set of vertices and $D, B, U$ denote sets of directed ($\diedgeright$), bidirected $(\biedge)$, and undirected $(\udedge)$ respectively. Within this class we implement methods and algorithms that are broadly applicable to any subclass: simple methods involving addition and deletion of edges, finding a subgraph $\G_S$ comprised  of only vertices in $S\subseteq V$ (and associated edges), and computing genealogical sets of a vertex, such as its ancestors, descendants, and siblings. %e.g., all ancestors of a vertex $V_i$ corresponding to all vertices $V_j$ such that a directed path $V_j \diedgeright \cdots \diedgeright V_i$ exists in $\G$.
We also implement a lightweight \texttt{draw} method using a Python interface to \texttt{graphviz} \citep{ellson2001graphviz, pygraphviz2022} for visualizing any instance of the class or subclasses of it -- all figures in this paper are produced using this functionality.
The rest of the inheritance structure is based on the types of edges each graph class contains. At the lowest levels are graphs only containing a single edge type: Directed acyclic graphs (DAGs) (only $\diedgeright$ edges) are the most popular type of causal graph \citep{robins1986new, spirtes2000causation, pearl2009causality}; Bidirected graphs (only $\biedge$ edges) are used to represent marginal correlations and are popular in genomics \citep{chaudhuri2007estimation, cox2014multivariate}; Undirected graphs (only $\udedge$ edges) can be used to encode feedback relationships \citep{lauritzen1996graphical}. Next, we have graphs containing a mixture of edges types: Acyclic directed mixed graphs (ADMGs) model systems with causal influence (via $\diedgeright$ edges) and correlation due to unmeasured confounding (via $\biedge$ edges) \citep{wright1921correlation, verma1990equivalence}; Chain graphs model causal influence (via $\diedgeright$ edges) as well as non-iid phenomena such as contagion, feedback, and symmetric relationships (via $\udedge$ edges) \citep{lauritzen2002chain, ogburn2020causal, bhattacharya2019causal}; Segregated graphs consisting of all three kinds of edges are capable of modeling all three mechanisms discussed above \citep{shpitser2015segregated}. We note that intrinsic graphs shown in the hierarchy of Fig.~\ref{fig:hierarchy} are not causal graphical models, but rather a graphical representation created by us to efficiently compute all statistical kernels required to parameterize a hidden variable causal model -- a necessary step for estimation  discussed in Section~\ref{sec:data_analysis}. This illustrates additional use cases of our graph inheritance structure for intermediate tasks. As another example, we use our chain graph implementation to encode equivalence classes of causal DAGs -- different models that imply the same restrictions on the observed data distribution -- known as Complete Partially Directed Acyclic Graphs (CPDAGs). This allows \texttt{Ananke} to easily interface with or extend causal discovery algorithms that output such objects, e.g., implementations of greedy equivalence search or the PC algorithm in the \texttt{causal-learn} package \citep{causallearn}.
\section{Data Analysis in \texttt{Ananke}}\label{sec:data_analysis}
\begin{figure}[H]
\begin{subfigure}{0.38\textwidth}
    \raggedright
 	\includegraphics[scale=.53]{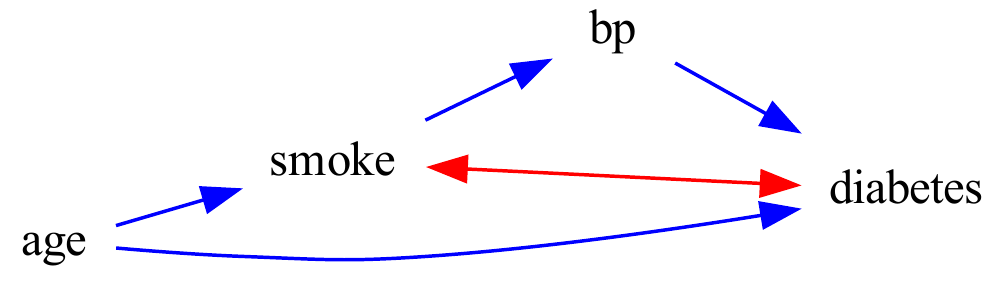} 
  
    \mbox{}
 	\end{subfigure}
\begin{subfigure}{0.62\textwidth}
\begin{minted}[fontsize=\footnotesize]{pycon}
>>> vertices = ["age", "smoke", "bp", "diabetes"]
>>> di_edges = [("age", "smoke"), ("smoke", "bp"),        
>>>             ("bp", "diabetes"), ("age", "diabetes")]
>>> bi_edges = [("smoke", "diabetes")]
>>> fdoor = graphs.ADMG(vertices, di_edges, bi_edges)
>>> fdoor.draw("LR")
\end{minted}
\end{subfigure}
\caption{Front-door model visualized using built-in capability.}
\label{fig:front-door-viz}
\end{figure}

To illustrate usage of \texttt{Ananke} we step through a hypothetical analysis for assessing the effect of smoking on diabetes  using a teaching dataset derived\footnote{The teaching extract of the Framingham Heart Study can be requested from \url{https://biolincc.nhlbi.nih.gov/teaching/.}} from the Framingham Heart Study \citep{kannel1968framingham}. %In this hypothetical analysis, we are interested in analyzing the effect of smoking on diabetes, as measured by glucose in the blood. 
We start by encoding substantive assumptions using an ADMG shown in \cref{fig:front-door-viz} along with the \texttt{Ananke} commands used to create and visualize it. %That is, we assume the effect of smoking on diabetes is mediated via blood pressure, age confounds smoking and diabetes, and other lifestyle choices that are not measured in the data confound smoking status and diabetes. %we also Finally, we know that lifestyle choices is a confounder between smoking status and diabetes, but it not measured in this study. Given the above, we may posit an ADMG, represented in . 

An ADMG can imply certain testable independence statements (amongst other general constraints) that can be read via m-separation \citep{richardson2003markov}. For example, we can verify that \texttt{age} is m-separated  from \texttt{bp} given \texttt{smoke} implying $\texttt{age}\ci \texttt{bp} \mid \texttt{smoke}$.
\begin{minted}[fontsize=\footnotesize]{pycon}
>>> fdoor.m_separated("age", "bp", ["smoke"])
True
\end{minted}
An analyst may verify whether the data supports such assumptions using any standard conditional independence test. Assuming Fig.~\ref{fig:front-door-viz} is correct, the next step is to apply identification theory to determine whether the desired causal effect can be expressed as a function of observed data. Applying \texttt{Ananke}'s implementation of a sound and complete algorithm for identification in presence of unmeasured confounding \citep{richardson2017nested} gives: 
\begin{minted}[fontsize=\footnotesize, escapeinside=||]{pycon}
>>> treatments = ["smoke"]; outcomes = ["diabetes"]
>>> id_fdoor = identification.OneLineID(graph=fdoor, treatments=treatments, outcomes=outcomes)
>>> print('Identifed =', id_fdoor.id(), '; Functional =', id_fdoor.functional()) 
Identifed = True 
|Functional = $\Sigma$agebp $\phi$diabetessmokebp(p(V);G) $\phi$smokediabetesage(p(V);G) $\phi$smokebpage(p(V);G)|
\end{minted}
That is, the counterfactual distribution $p(\texttt{diabetes}(\texttt{smoke}))$, and hence the effect, is indeed identified. Interpreting the output based on \cite{richardson2017nested} gives
{\footnotesize \[p(\texttt{diabetes}(\texttt{smoke})) = \sum_{\texttt{age}, \texttt{bp}}  p(\texttt{bp} \mid \texttt{smoke}) p(\texttt{age}) \left(\sum_{\texttt{smoke}'}p(\texttt{diabetes} \mid \texttt{smoke}', \texttt{bp}, \texttt{age})p(\texttt{smoke}' \mid \texttt{age} )\right).\]}
While the focus of this analysis is on identification under unmeasured confounding, \texttt{Ananke} also implements identification algorithms %asks if a particular causal effect is identified from a probability distribution given a set of assumptions encoded by a causal graph. While this can be done through application of the rules of do-calculus \citep{pearl1995causal}, advances in identification theory have produced sound and complete algorithms for identification in ADMGs \citep{shpitser2006identification, richardson2017nested}. 
for missing data \citep{rozi2019mid, nabi2020full}, selection bias, and data fusion \citep{lee2020general,lee2020identification}.
After identification, we may choose from a variety of estimation strategies offered in \texttt{Ananke}.

\subsection{Linear Gaussian Structural Equation Models}
One possible choice is to assume a linear structural equation model with correlated errors \citep{wright1934method}. In \texttt{Ananke}, we implement the iterative algorithm described in \cite{drton2009computing} to obtain maximum likelihood estimates for all edge coefficients; causal effects are then computed via path analysis. Applying this to standardized Framingham data gives:
%Linear Gaussian models structural equation models (SEMs) are popular for ADMGs in causal inference. Given a causal graph on $k$ variables, we have a directed edge coefficient matrix $B \in \mathbb{R}^{k \times k}$ and a positive definite covariance matrix for the error terms $\Omega \in \mathbb{R}^k \times k.$ Then, for the covariance matrix $\Sigma= (I-B)^{-T}\Omega(I-B)^{-1}$, the linear Gaussian SEM includes all normal distributions of the form ${\cal N} ({\bf 0}, \Sigma)$. Provided that the model is also bow-free, the residual iterative conditional fitting (RICF) algorithm proposed by  
%\cite{drton2009computing} may be used to learn $\Omega$ and $B$ via maximum likelihood.
\vspace{0.5em}
\begin{figure}[H]
\begin{subfigure}{0.45\textwidth}
    \centering
 	\includegraphics[scale=.4]{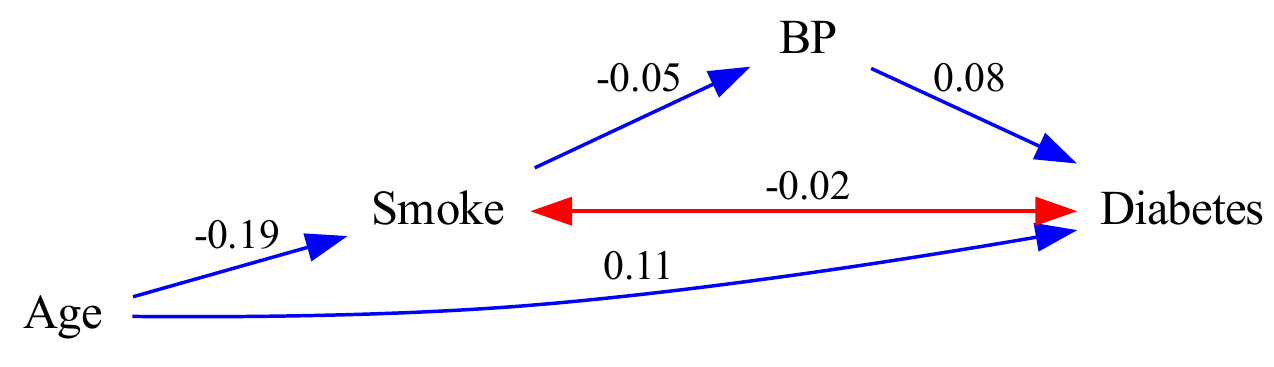}

    \vspace{-0.25cm}
    \mbox{}
\end{subfigure}
\begin{subfigure}{0.45\textwidth}
\vspace{-2em}
\begin{minted}[fontsize=\footnotesize]{pycon}
>>> lsm = LinearGaussianSEM(front_door)
>>> lsm = lsm.fit(df_cont)
>>> lsm.draw(direction="LR")
>>> lsm.total_effect(["smoke"], ["diabetes"])
ACE: -0.004
\end{minted}
\end{subfigure}
\end{figure}
\vspace{-2em}

\subsection{M\"{o}bius Parameterization for Discrete Data}
%\cite{evans2012maximum} proposed a smooth parameterization of the observed data likelihood for discrete ADMG models. However, the parameters are variationally independent and there is no closed form solution to the maximum likelihood estimation problem. %and method for fitting ADMGs to binary data via maximum likelihood. While estimation is relatively straightforward in DAGs, unobserved confounders in an ADMG render parameters of the M\"{o}bius parameterization  variationally dependent, and in general, there is no closed form solution. %, the complication of unobserved confounders requires a parameterization via the M\"{o}bius inverse transform. Because the parameters are not variationally independent, in general there is no closed form solution. However, the parameters can be learned via maximum likelihood through a coordinate descent algorithm proposed in \cite{evans2012maximum}.
An alternative is to use the M\"{o}bius parameterization of the observed data likelihood, which assumes all observed variables are discrete \citep{evans2012maximum}. We implement a coordinate descent algorithm to compute maximum likelihood estimates for the M\"{o}bius parameters (which may be variationally dependent in general.) Using binary versions of variables in the dataset, we obtain the following result for the average causal effect.
\begin{minted}[fontsize=\footnotesize]{pycon}
>>> bnm = binary_nested.BinaryNestedModel(front_door)
>>> bnm = bnm.fit(X=binary_nested.process_data(df_bin), tol=1e-12 )
>>> pY1_A0 = bnm.estimate(treatment_dict={"smoke": 0}, outcome_dict={"diabetes": 1})
>>> pY1_A1 = bnm.estimate(treatment_dict={"smoke": 1}, outcome_dict={"diabetes": 1})
>>> print("ACE: ", pY1_A1 - pY1_A0)
ACE: 0.004
\end{minted}

\subsection{Semiparametric Estimation of Causal Effects}
If the effect is identified,  \texttt{Ananke} lists several semiparametric estimation strategies, proposed by \cite{bhattacharya2022semiparametric}, and suggests the best one according to semiparametric efficiency theory. The implementation only requires specification of the ADMG, treatment, and outcome. 
%\cite{bhattacharya2020semiparametric} propose semiparametric estimators for a wide class of ADMGs. The implementation in \texttt{Ananke} only requires an ADMG and the target treatment and outcome -- if identified, \texttt{Ananke} suggests the best estimation strategy with respect to asymptotic efficiency. 
Using \texttt{Ananke}'s suggestion of \textit{efficient augmented primal IPW} estimator gives:
\begin{minted}[fontsize=\footnotesize]{pycon}
>>> ace_obj = CausalEffect(graph=front_door, treatment='smoke', outcome='diabetes')  
>>> ace = ace_obj.compute_effect(df_bin, "eff-apipw"); print("ACE: ", ace)
ACE: -0.002
\end{minted}

\acks{We thank Preethi Prakash and Ranjani Srinivasan for contributions to \texttt{Ananke}, and Carson Kurtz for assisting R.B. in testing code.}
% Acknowledgements should go at the end, before appendices and references

% \subsection{Visualizing Graphs}
% In \texttt{ananke}, a lightweight graph visualization capability that supports undirected, bidirected, and directed edges. This is done by using the \texttt{graphviz} package \citep{ellson2001graphviz} and the \texttt{pygraphviz} Python interface \citep{pygraphviz2022}.

% This capability is built in as a method of the parent class \texttt{graph}. For example, see \cref{fig:front-door-viz} as an example of the front-door model visualized using the built-in capability.

% Manual newpage inserted to improve layout of sample file - not
% needed in general before appendices/bibliography.

\bibliography{references}

\begin{thebibliography}{34}
\providecommand{\natexlab}[1]{#1}
\providecommand{\url}[1]{\texttt{#1}}
\expandafter\ifx\csname urlstyle\endcsname\relax
  \providecommand{\doi}[1]{doi: #1}\else
  \providecommand{\doi}{doi: \begingroup \urlstyle{rm}\Url}\fi

\bibitem[Bach et~al.(2022)Bach, Chernozhukov, Kurz, and
  Spindler]{DoubleML2022Python}
Philipp Bach, Victor Chernozhukov, Malte~S. Kurz, and Martin Spindler.
\newblock {DoubleML} -- {A}n object-oriented implementation of double machine
  learning in {P}ython.
\newblock \emph{Journal of Machine Learning Research}, 23\penalty0
  (53):\penalty0 1--6, 2022.
\newblock URL \url{http://jmlr.org/papers/v23/21-0862.html}.

\bibitem[Bhattacharya et~al.(2019{\natexlab{a}})Bhattacharya, Malinsky, and
  Shpitser]{bhattacharya2019causal}
Rohit Bhattacharya, Daniel Malinsky, and Ilya Shpitser.
\newblock Causal inference under interference and network uncertainty.
\newblock In \emph{Proceedings of the 35th Conference on Uncertainty in
  Artificial Intelligence}. AUAI Press, 2019{\natexlab{a}}.

\bibitem[Bhattacharya et~al.(2019{\natexlab{b}})Bhattacharya, Nabi, Shpitser,
  and Robins]{rozi2019mid}
Rohit Bhattacharya, Razieh Nabi, Ilya Shpitser, and James~M. Robins.
\newblock Identification in missing data models represented by directed acyclic
  graphs.
\newblock In \emph{Proceedings of the 35th Conference on Uncertainty in
  Artificial Intelligence}. AUAI Press, 2019{\natexlab{b}}.

\bibitem[Bhattacharya et~al.(2022)Bhattacharya, Nabi, and
  Shpitser]{bhattacharya2022semiparametric}
Rohit Bhattacharya, Razieh Nabi, and Ilya Shpitser.
\newblock Semiparametric inference for causal effects in graphical models with
  hidden variables.
\newblock \emph{Journal of Machine Learning Research}, 23:\penalty0 1--76,
  2022.

\bibitem[Chaudhuri et~al.(2007)Chaudhuri, Drton, and
  Richardson]{chaudhuri2007estimation}
Sanjay Chaudhuri, Mathias Drton, and Thomas~S Richardson.
\newblock Estimation of a covariance matrix with zeros.
\newblock \emph{Biometrika}, 94\penalty0 (1):\penalty0 199--216, 2007.

\bibitem[Cox and Wermuth(2014)]{cox2014multivariate}
David~Roxbee Cox and Nanny Wermuth.
\newblock \emph{Multivariate dependencies: Models, analysis and
  interpretation}.
\newblock Chapman and Hall/CRC, 2014.

\bibitem[Coyle(2021)]{coyle2021tmle3-rpkg}
Jeremy~R Coyle.
\newblock {tmle3}: The extensible {TMLE} framework.
\newblock \url{https://github.com/tlverse/tmle3}, 2021.
\newblock URL \url{https://doi.org/10.5281/zenodo.4603358}.
\newblock {R} package version 0.2.0.

\bibitem[Drton et~al.(2009)Drton, Eichler, and Richardson]{drton2009computing}
Mathias Drton, Michael Eichler, and Thomas~S. Richardson.
\newblock Computing maximum likelihood estimates in recursive linear models
  with correlated errors.
\newblock \emph{Journal of Machine Learning Research}, 10\penalty0 (10), 2009.

\bibitem[Ellson et~al.(2001)Ellson, Gansner, Koutsofios, North, and
  Woodhull]{ellson2001graphviz}
John Ellson, Emden Gansner, Lefteris Koutsofios, Stephen~C North, and Gordon
  Woodhull.
\newblock Graphviz—open source graph drawing tools.
\newblock In \emph{International Symposium on Graph Drawing}, pages 483--484.
  Springer, 2001.

\bibitem[Evans(2013)]{evans2013admgs}
Robin~J Evans.
\newblock {ADMG}s.
\newblock \url{https://www.stats.ox.ac.uk/~evans/software.htm}, 2013.

\bibitem[Evans and Richardson(2012)]{evans2012maximum}
Robin~J. Evans and Thomas~S. Richardson.
\newblock Maximum likelihood fitting of acyclic directed mixed graphs to binary
  data.
\newblock 2012.
\newblock URL \url{http://arxiv.org/abs/1203.3479}.

\bibitem[Hagberg et~al.(2022)Hagberg, Schult, and Renieris]{pygraphviz2022}
Aric Hagberg, Dan Schult, and Manos Renieris.
\newblock Pygraphviz.
\newblock 2022.
\newblock URL \url{https://github.com/pygraphviz/pygraphviz}.

\bibitem[Kalainathan et~al.(2020)Kalainathan, Goudet, and
  Dutta]{kalainathan2020causal}
Diviyan Kalainathan, Olivier Goudet, and Ritik Dutta.
\newblock Causal discovery toolbox: Uncovering causal relationships in python.
\newblock \emph{Journal of Machine Learning Research}, 21\penalty0
  (37):\penalty0 1--5, 2020.

\bibitem[Kalisch et~al.(2012)Kalisch, M{\"a}chler, Colombo, Maathuis, and
  B{\"u}hlmann]{kalisch2012causal}
Markus Kalisch, Martin M{\"a}chler, Diego Colombo, Marloes~H. Maathuis, and
  Peter B{\"u}hlmann.
\newblock Causal inference using graphical models with the {R} package pcalg.
\newblock \emph{Journal of Statistical Software}, 47:\penalty0 1--26, 2012.

\bibitem[Kannel and Gordon(1968)]{kannel1968framingham}
William~B. Kannel and Tavia Gordon.
\newblock \emph{The {F}ramingham Study: an epidemiological investigation of
  cardiovascular disease}.
\newblock Number 9-13. Department of Health, Education, and Welfare, 1968.

\bibitem[Kennedy(2021)]{kennedy2021npcausal}
Ed~Kennedy.
\newblock {npcausal}: Nonparametric causal inference methods.
\newblock \url{https://github.com/ehkennedy/npcausal}, 2021.

\bibitem[Lauritzen(1996)]{lauritzen1996graphical}
Steffen~L. Lauritzen.
\newblock \emph{Graphical Models}.
\newblock Oxford, U.K.: Clarendon, 1996.

\bibitem[Lauritzen and Richardson(2002)]{lauritzen2002chain}
Steffen~L. Lauritzen and Thomas~S. Richardson.
\newblock Chain graph models and their causal interpretations.
\newblock \emph{Journal of the Royal Statistical Society: Series B (Statistical
  Methodology)}, 64\penalty0 (3):\penalty0 321--348, 2002.

\bibitem[Lee and Shpitser(2020)]{lee2020identification}
Jaron J.~R. Lee and Ilya Shpitser.
\newblock Identification methods with arbitrary interventional distributions as
  inputs.
\newblock \emph{arXiv preprint arXiv:2004.01157}, 2020.

\bibitem[Lee et~al.(2020)Lee, Correa, and Bareinboim]{lee2020general}
Sanghack Lee, Juan~D Correa, and Elias Bareinboim.
\newblock General identifiability with arbitrary surrogate experiments.
\newblock In \emph{Uncertainty in artificial intelligence}, pages 389--398.
  PMLR, 2020.

\bibitem[Nabi et~al.(2020)Nabi, Bhattacharya, and Shpitser]{nabi2020full}
Razieh Nabi, Rohit Bhattacharya, and Ilya Shpitser.
\newblock Full law identification in graphical models of missing data:
  completeness results.
\newblock In \emph{International Conference on Machine Learning}, pages
  7153--7163. PMLR, 2020.

\bibitem[Ogburn et~al.(2020)Ogburn, Shpitser, and Lee]{ogburn2020causal}
Elizabeth~L. Ogburn, Ilya Shpitser, and Youjin Lee.
\newblock Causal inference, social networks and chain graphs.
\newblock \emph{Journal of the Royal Statistical Society: Series A (Statistics
  in Society)}, 183\penalty0 (4):\penalty0 1659--1676, 2020.

\bibitem[Pearl(2009)]{pearl2009causality}
Judea Pearl.
\newblock \emph{Causality}.
\newblock Cambridge University Press, 2009.

\bibitem[Richardson(2003)]{richardson2003markov}
Thomas~S. Richardson.
\newblock Markov properties for acyclic directed mixed graphs.
\newblock \emph{Scandinavian Journal of Statistics}, 30\penalty0 (1):\penalty0
  145--157, 2003.

\bibitem[Richardson et~al.(2017)Richardson, Evans, Robins, and
  Shpitser]{richardson2017nested}
Thomas~S. Richardson, Robin~J. Evans, James~M. Robins, and Ilya Shpitser.
\newblock Nested {M}arkov properties for acyclic directed mixed graphs, 2017.
\newblock {W}orking paper.

\bibitem[Robins(1986)]{robins1986new}
James~M. Robins.
\newblock A new approach to causal inference in mortality studies with a
  sustained exposure period -- application to control of the healthy worker
  survivor effect.
\newblock \emph{Mathematical Modelling}, 7\penalty0 (9-12):\penalty0
  1393--1512, 1986.

\bibitem[Scheines et~al.(1998)Scheines, Spirtes, Glymour, Meek, and
  Richardson]{scheines1998tetrad}
Richard Scheines, Peter Spirtes, Clark Glymour, Christopher Meek, and Thomas
  Richardson.
\newblock The tetrad project: Constraint based aids to causal model
  specification.
\newblock \emph{Multivariate Behavioral Research}, 33\penalty0 (1):\penalty0
  65--117, 1998.

\bibitem[Shpitser(2015)]{shpitser2015segregated}
Ilya Shpitser.
\newblock Segregated graphs and marginals of chain graph models.
\newblock In \emph{Advances in Neural Information Processing Systems}, pages
  1720--1728, 2015.

\bibitem[Spirtes et~al.(2000)Spirtes, Glymour, and
  Scheines]{spirtes2000causation}
Peter~L. Spirtes, Clark~N. Glymour, and Richard Scheines.
\newblock \emph{Causation, Prediction, and Search}.
\newblock MIT press, 2000.

\bibitem[Verma and Pearl(1990)]{verma1990equivalence}
Thomas Verma and Judea Pearl.
\newblock Equivalence and synthesis of causal models.
\newblock In \emph{Proceedings of the 6th Annual Conference on {U}ncertainty in
  {A}rtificial {I}ntelligence}, 1990.

\bibitem[Wright(1921)]{wright1921correlation}
Sewall Wright.
\newblock Correlation and causation.
\newblock \emph{Journal of Agricultural Research}, 20:\penalty0 557--580, 1921.

\bibitem[Wright(1934)]{wright1934method}
Sewall Wright.
\newblock The method of path coefficients.
\newblock \emph{Annals of Mathematical Statistics}, 5\penalty0 (3):\penalty0
  161--215, 1934.

\bibitem[Zevich(2018)]{zevich2018zepid}
Paul Zevich.
\newblock {zEpid}: An epidemiology analysis toolkit.
\newblock \url{https://github.com/pzivich/zepid/}, 2018.
\newblock Python package.

\bibitem[Zhang et~al.(2022)Zhang, Ramsey, Gong, Cai, Shimizu, Spirtes, Glymour,
  et~al.]{causallearn}
Kun Zhang, Joseph Ramsey, Mingming Gong, Ruichu Cai, Shohei Shimizu, Peter
  Spirtes, Clark Glymour, et~al.
\newblock causal-learn: {Causal Discovery for Python}.
\newblock https://github.com/cmu-phil/causal-learn, 2022.

\end{thebibliography}

\end{document}